\documentclass[preprint,preprintnumbers,amsmath,amssymb,superscriptaddress]{revtex4}
\usepackage{graphicx,color}
\usepackage{bm}
\usepackage{amsmath}
\usepackage{amsfonts}
\usepackage{amssymb}%
\usepackage{dcolumn}

\newcommand\Tm{\langle\mathbf{T}\rangle}

\newcommand{\be}{
\begin{equation}
}
\newcommand{\ee}{
\end{equation}
}
\newcommand{\beq}{
\begin{eqnarray}
}
\newcommand{\eeq}{
\end{eqnarray}
}
\begin{document}
\title{Mean first-passage time of an anisotropic diffusive searcher
 }

\author{N. Levernier}
\affiliation{Laboratoire de Physique Th\'eorique de la Mati\`ere Condens\'ee, CNRS/UPMC, 
 4 Place Jussieu, 75005 Paris, France}

\author{O. B\'enichou}
\affiliation{Laboratoire de Physique Th\'eorique de la Mati\`ere Condens\'ee, CNRS/UPMC, 
 4 Place Jussieu, 75005 Paris, France}
 
\author{R. Voituriez}
\affiliation{Laboratoire Jean Perrin, UMR 8237 CNRS /UPMC, 4 Place Jussieu, 75255
Paris Cedex, France}
\affiliation{Laboratoire de Physique Th\'eorique de la Mati\`ere Condens\'ee, CNRS/UPMC, 
 4 Place Jussieu, 75005 Paris, France}

\begin{abstract}
 We consider an anisotropic needle-like Brownian particle with nematic symmetry confined in a $2D$ domain. For this system,  the coupling  of translational and rotational diffusion makes the process ${\bf x} (t)$ of the positions of the particle  non Markovian. Using scaling arguments,  a Gaussian approximation and numerical methods, we determine the mean first passage time $\Tm$ of the particle to a target of radius $a$ and show in particular  that $\Tm\sim a^{-1/2}$ for $a\to 0$, in contrast with the classical logarithmic divergence obtained in the case of an isotropic $2D$ Brownian particle.
\end{abstract}

\maketitle

\section{Introduction}
Random search processes typically involve a randomly moving  searcher and a target -- for example a diffusive reactive particle and its reaction site or an animal looking for food \cite{Viswanathan:1999a,Benichou:2005qd,Shlesinger:2006,Benichou:2011fk,Oshanin:2007a,Rojo:2009}.   In this context, a useful observable to quantify the search efficiency  is  the mean first-passage time (MFPT) to the target \cite{Redner:2001a,bookSid2014,Singer:2006a,Schuss2007,Condamin2007,BenichouO.:2010,GuerinT.:2012fk,Guerin:2016qf,Benichou:2014fk,Benichou:2015bh,Bray:2013,Lucas:2014ty,Zhang:2016yq}. In the case of a single target in   a bounded domain -- or equivalently  regularly spaced targets in infinite space --, the full FPT statistics, and in particular its mean, have been derived asymptotically for Markovian scale invariant random walkers \cite{Condamin2007,BenichouO.:2010,Benichou:2014fk}. In the last few years these results have been  extended in several directions;  the case of several targets has been analyzed in \cite{Reingruber:2009,al:2011,Pillay:2010,Cheviakov:2010}, and more recently examples of non scale invariant processes  \cite{Tejedor:2012ly,Godec:2016,PhysRevLett.115.240601,Rupprecht:2016kq} and non Markovian processes have been studied. This was done first on the example of a monomer of a polymer chain \cite{GuerinT.:2012fk}, and next for more general Gaussian processes \cite{Guerin:2016qf}, but explicit results beyond the Markovian scale invariant case remain rather sparse. 

Here we consider the dynamics of an anisotropic (needle-like) particle in $2D$ (see Fig.\ref{fig1}); this system is of experimental relevance and can be realized by designing ellipsoidal particles or nematic macromolecules \cite{Han:2006fp,Ribrault:2007hl,Narayan:2007xe} and can model a large class of rod-like bacteria \cite{Doostmohammadi:2016ek}. On general grounds, the Brownian motion of such anisotropic particle, which could be of thermal or active origin,  combines rotational diffusion of the particle axis, and anisotropic translational diffusion along and orthogonal to the particle  axis. This coupling makes   the process ${\bf x} (t)$ of the positions of the particle  non Markovian, and not scale invariant, even if the full process $\left({\bf x} (t), \theta(t)\right)$, where $\theta$ denotes the orientation of the particle,  is Markovian. Adapting the recent methodology developed in \cite{GuerinT.:2012fk,Guerin:2016qf}, we determine in this paper the MFPT $\Tm$ of the particle to a target of radius $a$ and show in particular  that $\Tm\sim a^{-1/2}$ for $a\to 0$,  in contrast with the classical logarithmic divergence obtained in the case of an isotropic $2D$ Brownian particle, for which it is known that $\Tm\sim \log\left( r/a\right)$, where $r$ is the initial distance from the target  \cite{Redner:2001a,Condamin2007}. This classical results holds for both targets in the bulk of the domain, or at the boundary of the domain (so called narrow escape problem) \cite{Singer:2006a,Schuss2007,al:2011,Pillay:2010,Rojo:2012rw}. This logarithmic divergence  stems from the fact that the dimension of isotropic Brownian motion is 2. Exploration of the 2-dimensional plane by a Brownian particle has  henceforth specific properties, and is usually called marginally compact : in absence of confinement, any target of finite size $a\not=0$ is eventually visited with probability 1 (recurrence property).  In contrast,  in any dimension $D>2$, this probability is strictly smaller than 1 (transience property), and the MFPT in confinement diverges for small target size $a$;   in dimension $D<2$, a point-like target is found with probability 1 (recurrence property), and the MFPT has a finite limit when $a\to 0$ \cite{Hughes:1995}.

\section{General formulation of the mean-reaction time}

\begin{figure}[h]
\begin{center}
\includegraphics[scale=0.43]{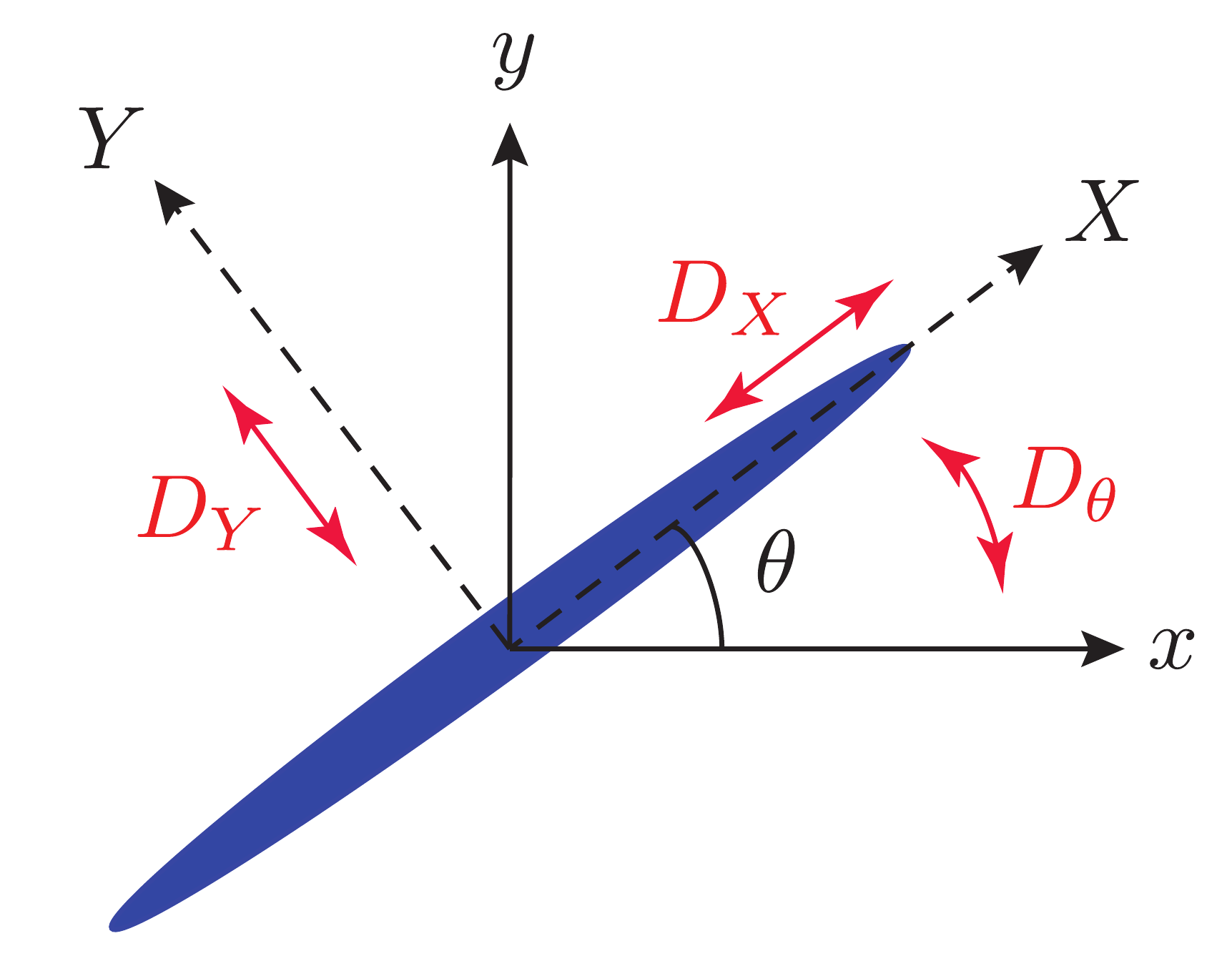} 
\end{center}
\caption{Local coordinate system and definition of parameters for an anisotropic Brownian particle.}
\label{fig1}
\end{figure}

The 2D diffusion of anisotropic particles, such as ellipsoid particles or more generally nematic particles, can be conveniently defined in the local reference frame of the particle, where the  $X$ axis is defined by  the main axis of the particle at a given time $t'$ and  the $Y$ axis by its normal.  The position and state of the particle at a later time $t>t'$ is then fully defined by $(X,Y,\theta)$, where $X,Y$  refer to the position of the center of mass of the particle in the local reference frame and the angle $\theta$ defines its orientation  with respect to a fixed axis (see Fig. \ref{fig1}). 
The dynamics can then be written:
\begin{equation}
\partial_t X=\xi_X;\  \partial_t Y=\xi_Y;\  \partial_t \theta=\xi_{\theta} 
\end{equation}
The noise terms $\xi_i$ are here assumed to be Gaussian and characterized by $\langle \xi_{\theta}(t) \xi_{\theta}(t')  \rangle = 2 D_{\theta} \delta (t-t')$ and $\langle \xi_{i}(t) \xi_{j}(t')  \rangle = 2 \delta_{ij} D_{i} \delta (t-t')$ and $\langle \xi_{\theta}(t) \xi_{j}(t')  \rangle = 0$  for $i,j\in\{X,Y\}$.
 In a fixed reference frame, the dynamics of the position  ${\bf x} (t)$ of the particle can then be derived and reads in cartesian coordinates
\begin{equation}
\partial_t x=\zeta_x;\  \partial_t y=\zeta_y;\  \partial_t \theta=\xi_{\theta}.
\end{equation}
In this fixed reference frame the noise terms now have non trivial correlations defined by $\langle \zeta_{i}(t) \zeta_{j}(t')  \rangle = 2 \mu_{ij} \delta (t-t')$ for $i,j\in\{x,y\}$, where the correlation matrix   $\mu$ is given by :
\begin{equation}
\mu=\frac{D_{X}+D_{Y}}{2} \,{I}+\frac{D_{X}-D_{Y}}{2}\, M_{\theta}
\end{equation}
with
\begin{equation}
M_{\theta}=\begin{pmatrix}
   \cos{2 \theta} &  \sin{2 \theta}\\
   \sin{2 \theta} & -\cos{2 \theta} 
\end{pmatrix}
\end{equation}
and $I$ stands for the identity matrix.
Note that the angular variable $\theta$ is  Brownian with  diffusion coefficient $D_{\theta}$, and is therefore  Gaussian, and that the full process $\left({\bf x} (t),\theta(t)\right)$ is Markovian. The non trivial correlation matrix $\mu$ shows that the process ${\bf x} (t)$ alone is non Markovian and \emph{a priori} not Gaussian. It is however useful to determine the diffusion tensor for a given initial angle $\theta_{0}$, which can be written \cite{Han:2006fp} :
\begin{equation}
\frac{\left \langle \Delta x_{i} (t) \Delta x_{j} (t) \right\rangle }{ 2t} = \frac{D_{X}+D_{Y}}{2} \,\delta_{ij}+\frac{D_{X}-D_{Y}}{2}\,g(D_{\theta} t)\, \left(M_{\theta_{0}}\right)_{ij}
\label{diffTensor}
\end{equation}
with $g(x)=(1-e^{-4x})/(4x)$.

\begin{figure}[h]
\begin{center}
\includegraphics[scale=0.43]{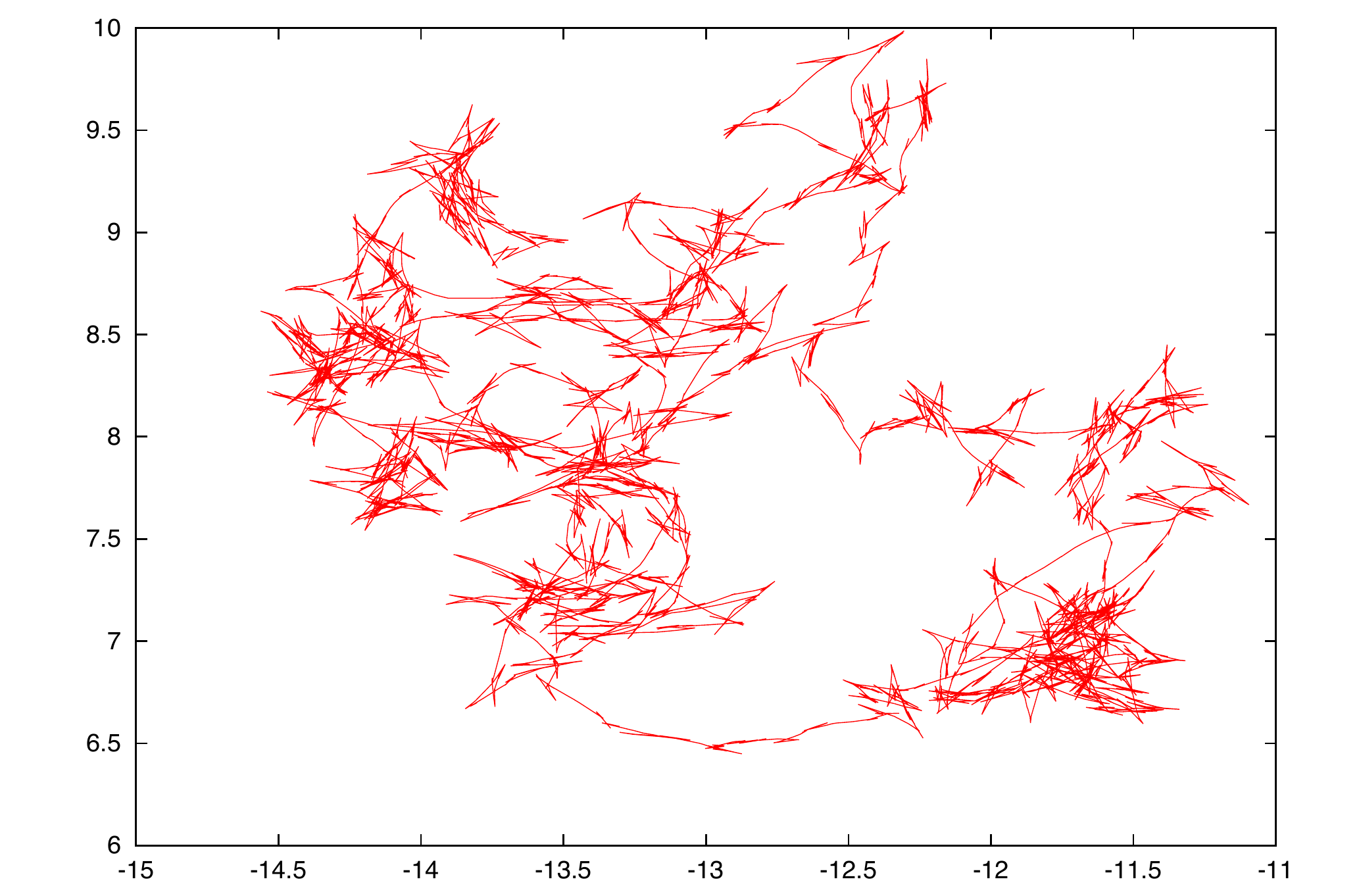} 
\end{center}
\caption{Example of Brownian trajectory of a needle-like anisotropic particle ($D_Y=0$) with $D$ and $D_{\theta}$ arbitrarily set to 1.}
\label{fig2}
\end{figure}

Taking $\theta_{0}=0$, this equation has a clear interpretation.  At small time scales (compared to the characteristic time scale of rotational diffusion $1/D_\theta$), each coordinate $x$ and $y$ diffuse  independently with  diffusion coefficients $D_{X}$ and $D_{Y}$; at  time scales of the order $1/D_\theta$,  $x$ and $y$ are non trivially coupled according to Eq. (\ref{diffTensor}); finally at time scales larger than $1/D_\theta$  both $x$ and $y$  diffuse independently with the averaged diffusion coefficient $(D_{X}+D_{Y})/2$ and $2D$ Brownian motion is recovered (see Fig.\ref{fig2} for a typical trajectory) .

In what follows, we investigate the mean reaction time of such anisotropic particle  with a spherical target $\cal T$ of radius $a$, located in a $2D$ domain of volume $V$. Here the reaction time is defined as the first-passage time of the particle center of mass at the target. The particle is assumed to start at a distance $R_{0}$ from the center of the target with a uniformly distributed orientation $\theta_0$. As detailed below we will consider the large $V$ limit, defined by taking all points of the boundary to infinity while the target and starting point remain fix. In this limit the specific nature of boundary conditions (for example reflecting) is irrelevant, and the dependence on the starting position can occur only through the distance $R_0$; we will therefore assume that the initial position is uniformly distributed on the circle of radius $R_0$. Unless specified otherwise, we will also consider the regime $R_0\gg a$.
Last, to highlight the effect of anisotropic diffusion, we will assume below that $D_{X}=D$ and $D_{Y}=0$.

The starting point of our analysis is to use the fact that the full process $\left({\bf x} (t),\theta(t)\right)$ is Markovian. A renewal equation can therefore be written for a given ${\bf x}\in{\cal T}$:
\begin{equation}
P({\bf x},\theta,t|{\rm ini})=\int_{0}^{t} dt'\int_{{\bf x}'\in \partial {\cal T}} d{\bf x}'  \int_0^\pi d\theta'  F({\bf x}',\theta',t'|{\rm ini})P({\bf x},\theta,t|{\bf x}',\theta',t') 
\label{renewal}
\end{equation}
where $P({\bf x},\theta,t|{\bf x}',\theta',t')$ denotes the full propagator, namely the probability density of finding the anisotropic particle at $({\bf x},\theta)$ at time $t$ knowing that it was at $({\bf x}',\theta')$ at $t'$. Similarly, $F({\bf x},\theta,t|{\bf x}',\theta',t')$ denotes the probability density that the particle hits the target for the first time at  $({\bf x},\theta)$ at time $t$ knowing that it started at $({\bf x}',\theta')$ at $t'$. Here ini stands for the initial conditions at $t=0$ as given above. Thanks to the rotational symmetry of the initial condition, $F$ does not depend on $x'$, and the integration over $\partial \cal T$ is  trivial. Following \cite{GuerinT.:2012fk}, we next introduce the distribution $\pi(\theta)$, which gives the probability density of  the orientation angle $\theta$ at the instant of reaction ($\pi(\theta)=\int F(\theta,t'|{\rm ini}) dt'$).  Taking the Laplace transform of Eq.(\ref{renewal}) then leads to the following expression of the MFPT $\Tm$ (see \cite{GuerinT.:2012fk,Guerin:2013zr,Guerin:2013ly}):
\begin{equation}
\frac{\Tm}{V}=\int_{0}^{\infty} dt  \left[\bar  P({\bf x},t|\pi) - \bar P({\bf x},t|{\rm ini})\right]
\label{eqT}
\end{equation}
where we have defined 
\begin{equation}
\bar P({\bf x},t|{\rm ini}) =\int_0^\pi d\theta  P({\bf x},\theta,t|{\rm ini}) 
\label{Px}
\end{equation}
and 
\begin{equation}
\bar P({\bf x},t|\pi) =\int_{{\bf x}'\in \partial {\cal T}} d{\bf x}'  \int_0^\pi d\theta'  P({\bf x},t|{\bf x}',\theta',t'=0) \pi(\theta')
\label{Ppi}
\end{equation}
Equations (\ref{eqT}) and (\ref{Ppi}) together with the normalization condition for $\pi(\theta)$  yield an integral equation that fully defines $\pi$ and $\Tm$. Note that this definition holds for any point ${\bf x}\in\cal T$. At this stage,  for a finite confining domain, the propagator $\bar P$ depends on the full geometry of the problem and cannot be determined explicitly. We will therefore consider below the $V\to\infty $ limit of equation (\ref{eqT}) and still denote $\bar P$ the propagators in infinite space for the sake of simplicity. It should be mentioned that even in infinite space there is no known analytical expression for the propagator $\bar P$, which excludes an explicit resolution of this integral system.
We however show below that a useful approximate scheme can be developed and that a scaling analysis can be performed. As discussed in the next section, it will be useful to introduce the characteristic length $\delta\equiv\sqrt{D/D_\theta}$; at length scale smaller than $\delta$ the motion of the particle is anisotropic, while isotropic diffusion is recovered at length scales larger than $\delta$. One can then define the dimensionless parameter $\beta\equiv (a/\delta)^2$: for $\beta \ll 1$ angular diffusion is slow at the scale of the target, so that orientation is conserved when translational diffusion covers a typical distance $a$ and conversely for  $\beta \gg 1$ angular diffusion is fast.

\section{Gaussian approximation }

As mentioned above, it is known  that even in infinite space the propagator $\bar P$ of the process ${\bf x}$ alone is not exactly Gaussian \cite{Ribrault:2007hl}. However, as checked numerically below, taking a Gaussian approximation of this propagator, with exact mean and covariance leads to an accurate determination of the MFPT $\Tm$.
Within this approximation, in a coordinate system where the center $C$ of the target is at $(0,0)$, and choosing ${\bf x}=C$ in Eq.(\ref{eqT}), the propagators entering Eq.(\ref{eqT}) can then be expressed in terms of :
\begin{equation}
\bar P(0,0,t | x,0,\theta_{0},0)=\frac{1}{2\pi D t \sqrt{1-g^{2}(D_{\theta}t)}} \exp \left(-\frac{x^{2}}{2Dt} \,\frac{1-g(D_{\theta}t)\cos 2\theta_{0}}{1-g^{2}(D_{\theta}t)} \right).
\end{equation}
The distribution $\pi(\theta)$ then remains to be determined. In general, this distribution seems difficult to derive explicitly;  useful limiting regimes can however be determined as follows. (i) $\beta \gg 1$. In this regime angular diffusion is very fast at the scale of the target and $\pi (\theta)$ is uniform. This is the Brownian limit. 
(ii) $\beta \ll 1 $. Angular diffusion is  slow at the scale of the target, and simple geometric arguments yields  $\pi (\theta) = \cos (\theta) /2$ in the regime $R_0\gg a$.  
We focus from now on on the  case $\beta \ll 1$,  for which the effect of anisotropy is expected to be important. After performing the integration over the angle $\theta $ at the instant of reaction, one gets from Eq.(\ref{eqT}) :
\begin{align}
\lim_{V\to\infty}\frac{\Tm}{V}=\int_{0}^{\infty} du  &\left[\frac{1}{4 D \sqrt{\pi \beta u g(u)}}\exp \left(-\frac{\beta}{2u(1+g(u))}  \right) \text{Erf}\left(  \sqrt{\frac{\beta}{u}}\sqrt{\frac{g(u)}{1-g^{2}(u)}} \right) \right. \nonumber \\
& \left.- \frac{1}{2\pi D u \sqrt{1-g^{2}(u)}}\exp \left(-\frac{\gamma}{2u} \frac{1}{1-g^{2}(u)} \right) I_0\left( 0, \frac{\gamma}{2u}\, \frac{g(u)}{1-g^{2}(u)} \right)  \right]\nonumber \\
&\equiv  \int_{0}^{\infty} du [H_1(\gamma,\beta,u)-H_2(\gamma,u)]\label{MFPT}
\end{align}
where $\gamma=R_{0}^{2}D_{\theta}/D=\beta R_{0}^{2}/a^{2}$ and $H_1,H_2$ are defined by obvious identification with the first two lines of Eq.(\ref{MFPT}). Here $I_0$ denotes the modified Bessel function. Two relevant asymptotic regimes can then be analyzed.

\emph{Small target size limit:  $\beta \rightarrow 0$ and fixed $\gamma$. } We first focus on the dependence of the MFPT on the target size $a$ and analyze the regime of small $a$, which amounts to taking $\beta\to 0$. First note that the integral giving $\Tm$ in Eq.(\ref{MFPT}) diverges for $\beta \rightarrow 0$, which shows that the MFPT diverges in the small target size limit, as is the case for regular Brownian motion in space dimension $D\ge2$.
 This divergence is controlled by the behaviour of the propagators at small time scales. It is  therefore useful to introduce the intermediate scale $\epsilon$ such that  $\beta\ll \epsilon \ll 1$. Making use of $g(x)=1-2x+o(x)$ and $I_0(x) \sim e^{x}/\sqrt{2\pi x}$ for $x\to\infty$,  one can rewrite :
\begin{align}
\lim_{V\to\infty}\frac{\Tm}{V}=& \frac{1}{4 D \sqrt{\pi \beta}} \int_{0}^{\epsilon} du \frac{1}{\sqrt{u}} \exp \left( -\frac{\beta}{4u}\right) \text{Erf}\left(\frac{\sqrt{\beta}}{2u}\right)  
-\frac{1}{4 D \sqrt{\pi \gamma}} \int_{0}^{\epsilon} du \frac{1}{\sqrt{u}} \exp \left( -\frac{\gamma}{4}(1+1/u)\right) \nonumber \\
& + \int_{\epsilon}^{\infty} du [H_1(\gamma,\beta,u)-H_2(\gamma,u)]. \label{MFPT2}
\end{align}
The $\beta\to 0$ limit can then be taken and yields finally
\begin{equation}
\lim_{V\to\infty}\frac{\Tm}{V}\sim  \beta^{-1/4}\times \frac{1}{4 D \sqrt{2\pi}} \int_{0}^{\infty} du \frac{\text{Erf}(u)}{u^{3/2}}.\label{pref}
\end{equation}
This shows in particular that in this small target regime, the MFPT diverges as $\Tm\sim a^{-1/2}$, which stands in contrast with classical logarithmic divergence obtained in the Brownian $2D$ case.

\emph{Large initial distance limit : $\gamma \rightarrow \infty$ and fixed $\beta$ }. We now focus on the dependence on $R_0$, for $R_0$ large, which amounts to taking the large $\gamma$ asymptotics in Eq.(\ref{MFPT}). Making use again of the decomposition given in Eq.(\ref{MFPT2}), one obtains straightforwardly
\begin{equation}
\lim_{V\to\infty}\frac{\Tm}{V}\sim\underset{\gamma \rightarrow \infty}\sim \frac{1}{2 \pi D} \log \gamma.
\end{equation}
Rewriting $\gamma=(R_0/\tilde a)^2$, where we introduce $\tilde a=\sqrt{D/D_\theta}$, one recovers the MFPT of a $2D$ isotropic Brownian motion of diffusion coefficient $D/2$ (which is indeed the large scale diffusion coefficient of the particle) to a spherical target of radius $\tilde a$.

\section{Asymptotic scaling analysis}

We now present a more general scaling analysis of the problem, which does not involve the above Gaussian approximation of the  propagator. As discussed above, one needs to distinguish the two time regimes   $t \ll 1/D_{\theta}$ of highly anisotropic  diffusion and $t \gg 1/D_{\theta}$ of isotropic diffusion. 
A simple dimensional analysis shows that the rescaled MFPT $\tau\equiv D\Tm/V$ can be written $\tau=f(a/\delta,R_{0}/\delta,L/\delta)$, where   $f$ is a dimensionless function,  $L=V^{1/2}$ is the characteristic domain size, and $\delta=\sqrt{D/D_\theta}$ has been defined above. First notice that the existence of a finite limit of $\Tm/V$ when $V\to\infty$, as can be seen from  Eq.(\ref{eqT}), shows that the dependence on the  last variable $L/\delta$ can be ignored; we consider this limit from now on and identify $\tau$ and its large volume limit.

\begin{figure}[h]
\begin{center}
\includegraphics[scale=0.63]{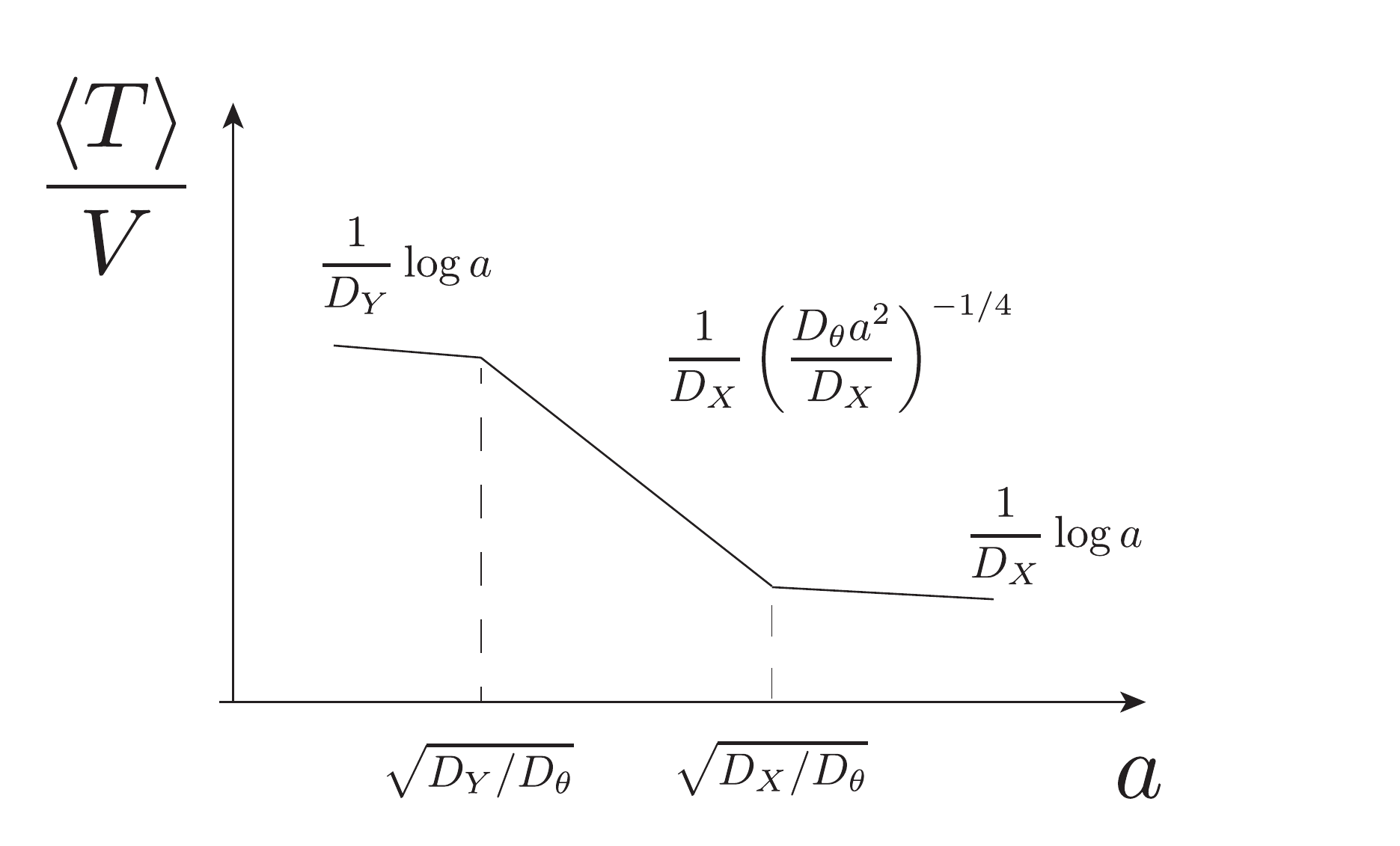} 
\end{center}
\caption{Sketch of the dependance of the MFPT $\Tm$ on the target size $a$ for $D_{Y}\ll D_{X}$.}
\label{sketch}
\end{figure}

We analyze the dependence on the target size $a$ and fix $R_0/\delta$; the rescaled MFPT $\tau$ is then a function of $a/\delta$ only. This function is expected to diverge for $a/\delta\to 0$; qualitatively  this comes from the fact that the anisotropic motion results from a constrained $2D$ Brownian motion and can therefore only explore less space (see below for a more quantitative argument). From Eq. \eqref{eqT} such  divergence comes from the contribution of the first propagator $\bar P({\bf x},t|\pi)$, which only depends on $a$. Dimensional analysis shows that one can write  $P({\bf x},t|\pi)=g(a/\sigma(t))/v(t)$, where $\sigma(t)$ is a length, $v(t)$ a surface area and $g$ a dimensionless function. The analysis of the diffusion tensor given in Eq. \eqref{diffTensor} shows that at short time scales one has to define two distinct characteristic lengths $\sigma_{x}=\sqrt{Dt}$ and  $\sigma_{y}=\sqrt{D D_{\theta}} t$. We argue here that at short time scales the characteristic surface area should be defined by $v(t)\propto \sigma_{x}\sigma_{y}\propto D \sqrt{D_{\theta	}} t^{3/2}$, which gives the characteristic surface area covered by the trajectory at short times. In turn, $\sigma(t)$ should be taken as the smallest length scale, and we therefore anticipate that $\sigma\propto\sigma_{y}$. Finally, injecting these scalings in  Eq.(\ref{eqT}), one obtains :
\begin{align}
\tau &\sim \int \frac{f(a/\sqrt{D D_{\theta}} t)}{  \sqrt{D_{\theta}} t^{3/2} }dt \\
& \sim \sqrt{\frac{\delta}{a}} \int \frac{f(1/u)}{u^{3/2}}du.
\end{align}
This analysis, even if it involves assumptions on the scaling of the propagator, is consistent with the analysis of the previous section based on the Gaussian approximation of the propagator and yields the correct scaling $\tau\sim \beta^{-1/4}$; it will be further supported in the next section by numerical simulations.
To conclude this general analysis, we extend it to the case of a finite transverse diffusion coefficient $D_Y\not=0$, while keeping   $D_{Y} \ll D_X$. Three cases then arise. For $D_{Y}/D_{\theta} \ll a^{2} \ll D_{X}/D_{\theta}$, the diffusion along $Y$ can be neglected and the previous analysis can be applied : $\Tm \propto \beta^{-1/4}$. In the ultimate regime $D_{Y}/D_{\theta} \gg a^{2}$, the particle behaves as $2D$ Brownian particle, and the usual logarithmic divergence is recovered. Diffusion is however inhomogeneous and the FPT is limited by the transverse motion, so that  $\Tm \propto (\log{a}/R_0)/D_Y$. In the opposite regime $D_{X}/D_{\theta} \ll a^{2}$, the  kinetics are limited by the large scale motion, which is diffusive with diffusion coefficient $D_X/2$ so that  $\Tm \propto (\log{a}/R_0)/D_X$ (see Fig. \ref{sketch}).

\section{Simulations and refined approximations}

\begin{figure}[h]
\begin{center}
\includegraphics[scale=0.8]{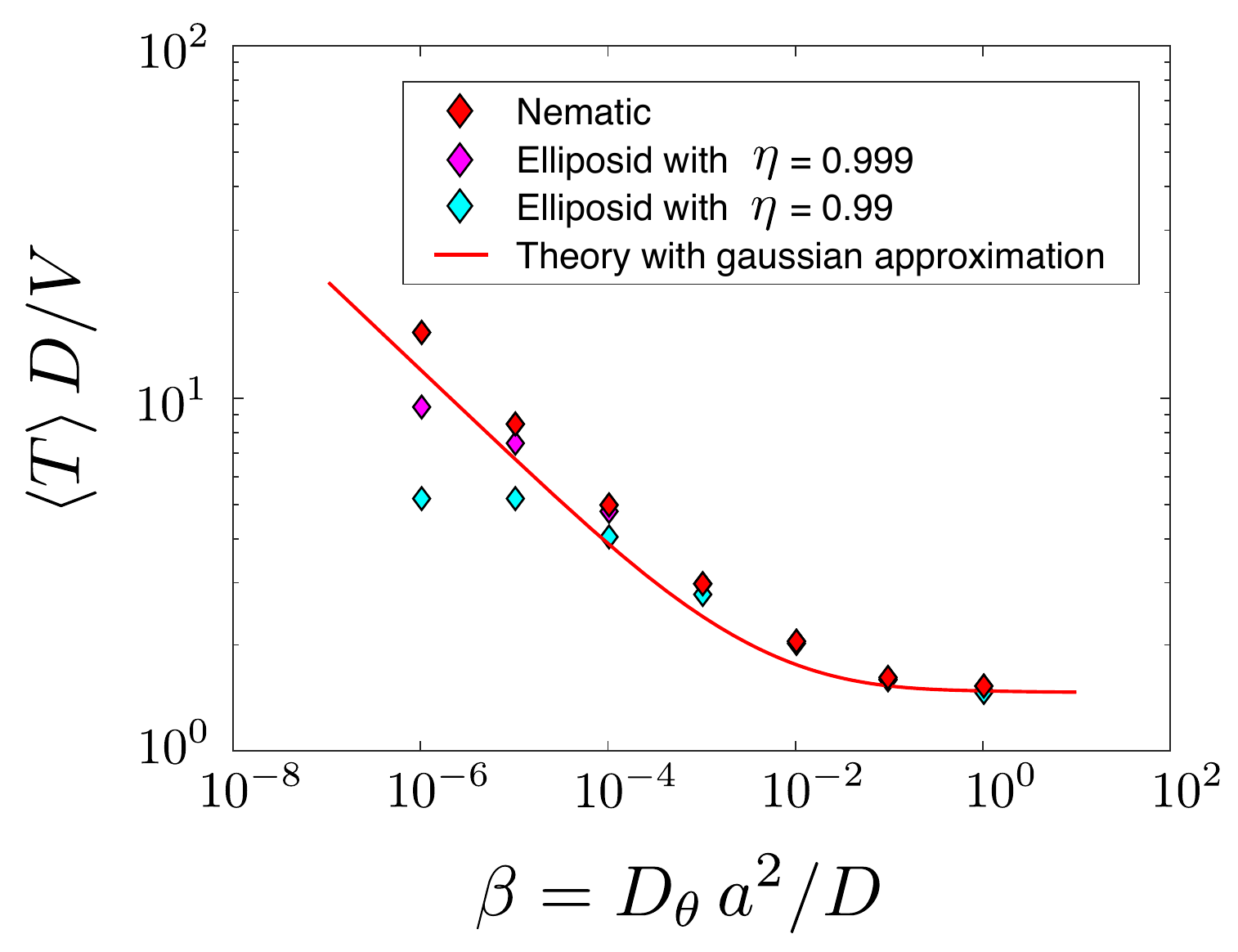} 
\end{center}
\caption{MFPT of an anisotropic diffusing particle to a target as a function of $\beta$ for different values of the anisotropy coefficient $\eta=(D_X-D_Y)/(D_X+D_Y)$. Theoretical predictions (plain lines) and numerical simulations (symbols) are shown. Parameters : $D=1$, $a=1$, $X_0=100$, $V=\pi\times 500^{2}$. The scaling law $\beta^{-1/4}$ is observed as predicted, although the numerical prefactor given by the gaussian approximation is underestimated by about 17\%.}
\label{fig3}
\end{figure}

We now compare our results with numerical simulations, based on a  simple discretization of Langevin equation as suggested in \cite{Ribrault:2007hl}. Fig. \ref{fig3} shows  the MFPT as a function of $\beta$. The predicted scaling $\sim\beta^{-1/4}$ for small $\beta$ is clearly observed numerically, with a cross over to the isotropic Brownian behavior at larger $\beta$ in the case $D_Y=0$. We present additionally several examples with $D_Y\not=0$, keeping  $D_{Y}/D_{X}\ll 1$, which display as expected a similar behavior. 
However, it should be noted that the numerical prefactor entering the expression of the MFPT, as obtained in Eq. (\ref{pref}) on the basis of the Gaussian approximation, is not quantitatively accurate. This is   due to the fact that the propagator is not exactly Gaussian, as can be checked numerically.  The good agreement that we obtain for the scaling with $\beta$ however suggests that our scaling hypothesis, consistent with the Gaussian approximation of the propagator, is correct.

\section{Conclusion}

In conclusion, we have presented in this paper a theoretical analysis of the MFPT to spherical target for  an  anisotropic diffusive particle in a $2D$ confined domain. On the basis of scaling arguments,  a Gaussian approximation and numerical methods, we have determined the MFPT $\Tm$ and put forward an unexpected scaling with the target size   $\Tm\sim a^{-1/2}$ for $a\to 0$, which is in contrast with the classical logarithmic divergence obtained in the case of an isotropic $2D$ Brownian particle. This stronger divergence of the MFPT for small target sizes can be attributed to a much slower exploration of space at short time scales for the anisotropic particle : as discussed above, at short times ($t<1/D_\theta$) the anisotropic particle explores a typical area $\sim t^{3/2}$, much smaller than the area $\sim t$ explored by a $2D$ isotropic Brownian particle. Since it is expected that the distribution of the first-passage time is in this case asymptotically exponential, we anticipate that this quantity allows for a complete description  of the search kinetics \cite{BenichouO.:2010}.

 It should be noted that this non trivial scaling of the MFPT with the target size cannot be directly inferred from the general results derived in \cite{Condamin2007} for scale invariant processes; the process is here not scale invariant and a single walk dimension cannot a priori be defined. While this work focuses on a rather  simple model of anisotropic diffusion, we believe that it also puts forward tools applicable to further examples of anisotropic  processes, which lack the scale invariance property. It also provides an explicit example of determination of the MFPT for a non Markovian process ($\bf x$ alone), which is amenable to a Markovian process by taking into account additional degrees of freedom (${\bf x}, \theta$). In particular this analysis illustrates that the specific functional form of the propagator can be irrelevant  to determine scaling behaviors, provided that relevant rescaled variables are properly defined.


\end{document}